\documentclass[prl,twocolumn,aps,superscriptaddress,showpacs]{revtex4}
\usepackage{graphicx}
\usepackage{amsmath}
\usepackage{amssymb}
\usepackage{tabularx}
\usepackage{multirow}
 
\newcommand{\PreserveBackslash}[1]{\let\temp=\\#1\let\\=\temp}

\begin{document}       

\title{ Finding the center reliably: robust patterns of developmental
gene expression}

\author{
Martin Howard}
\affiliation{Department of Mathematics, Imperial College London, South
  Kensington Campus,  
London SW7 2AZ, UK.}
\author{Pieter Rein ten Wolde}
\affiliation{FOM Institute for Atomic and Molecular Physics (AMOLF),
  Kruislaan 407, 1098 SJ, Amsterdam, The Netherlands.}

\date{\today}
\begin{abstract}
We investigate a mechanism for the robust identification of the center
of a developing biological system.  We assume the existence of two
morphogen gradients, an activator emanating from the anterior, and a
co-repressor from the posterior. The co-repressor inhibits the
action of the activator in switching on target genes. We apply this
system to {\em Drosophila} embryos, where we predict the existence of
a hitherto undetected posterior co-repressor. Using
mathematical modelling, we show that a symmetric
activator-co-repressor model can quantitatively explain the precise
mid-embryo expression boundary of the {\em hunchback} gene, and the
scaling of this pattern with embryo size.
\end{abstract}


\maketitle 

During embryonic development, 
cells have to differentiate in a manner
that is dictated by their positions within the developing
embryo. Turing conjectured that the positional information for
differentiating cells is provided by molecules called morphogens that
self-organize into spatial patterns~\cite{Turing52}. Today a number of
proteins have been identified that act as morphogens. However, in
contrast to Turing's original conjecture, morphogens are usually
produced at a localized source, after which they diffuse into the
surrounding tissue. Here, morphogens are also degraded and a
concentration gradient is thus formed. The differentiating cells
respond via patterns of gene expression that depend on the morphogen
concentration~\cite{Wolpert69}.  However, the biochemical processes
that give rise to the morphogen gradient are prone to variations: the
rates of protein synthesis/degradation, for example, will vary from
embryo to embryo. Variation in these parameters might modify the
morphogen gradient and thereby induce error in the positional
information transmitted to the differentiating cells. Nevertheless,
cell differentiation is often exceedingly
precise~\cite{Houchmandzadeh02}. How embryonic development is robust
against fluctuations in the underlying biochemical processes is still
poorly understood.

Recently, several mechanisms have been proposed to explain
robust embryonic development~\cite{Eldar02,Eldar03,Bollenbach05}. They
rely on
 the insensitivity of
the morphogen gradient to embryo-to-embryo variations in the morphogen
synthesis rates~\cite{Eldar02,Eldar03,Bollenbach05}. However, these
mechanisms cannot explain recent experimental observations on the
Bicoid-Hunchback system in the developing {\em Drosophila}
embryo~\cite{Houchmandzadeh02}. {\em bicoid} ({\em bcd}) messenger RNA
(mRNA) is deposited at the anterior pole of the embryo, where it forms
a localized source of Bcd protein. The Bcd protein diffuses away from
the pole and is degraded, thereby forming a concentration gradient
(see Fig.1a). A simple `threshold' model would postulate that
downstream genes such as {\em hunchback} ({\em hb}) are activated only
when the concentration of Bcd is above a certain
threshold~\cite{Wolpert69}. However, while the {\em hb} expression
domain is highly precise, the Bcd morphogen gradient exhibits large
embryo-to-embryo variations~\cite{Houchmandzadeh02, Spirov03}. This
rules out the above mechanisms, which rely on the robust formation of
the morphogen gradient itself. Moreover, experiments reveal that the
Hb boundary, $x_{\rm Hb}$, scales with the embryo
length (EL): $x_{\rm Hb} = 0.49 \, (\pm \, 0.01)$ EL~\cite{Houchmandzadeh02}.
Such scaling of
expression domains cannot be explained by a single morphogen-gradient
model, in which positional information is only emanating from one
end. Indeed, this observation suggests that the formation of the {\em
hb} expression domain is regulated by two or more morphogens, sourced
at opposite ends of the embryo. Earlier work of
Wolpert~\cite{Wolpert69} and Meinhardt~\cite{Meinhardt98} already
suggested that such models might provide size regulation.

\begin{figure}[b] \centering
\includegraphics[width=8cm]{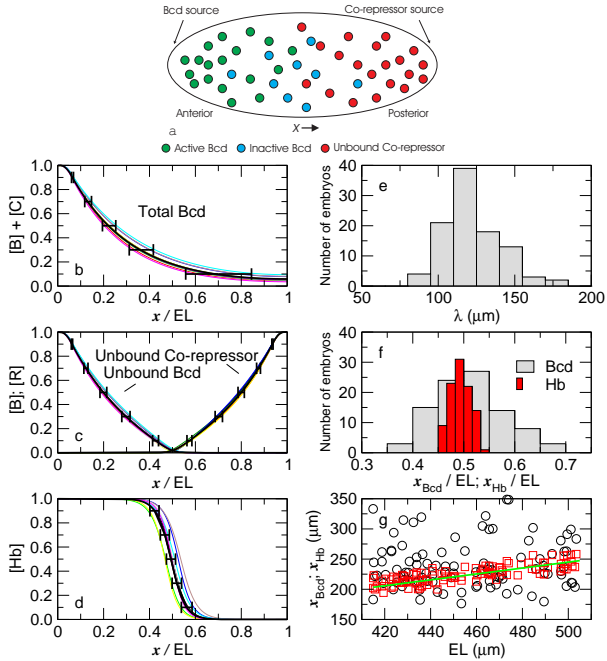} \caption{\label{fig:1}
a: Sketch of {\em Drosophila} embryo showing localized
Bcd/co-repressor sources at opposite poles, and co-repressor-Bcd
binding. b-g: Positional information for Bcd and Hb density profiles;
data from 100 simulated embryos.  Maximum densities normalized to
unity. Mean density profile and standard deviation for: b: Total Bcd (Bcd
plus complex); c: Unbound Bcd and co-repressor; d: Hb; in b)-d), $10$
typical individual density profiles are also shown. e: Distribution of
length scale $\lambda$ describing exponential decay of total Bcd
density profile. f: Distribution of positions $x_{\rm Bcd}$ and
$x_{\rm Hb}$ where each Bcd and Hb density profile crosses given
threshold, 0.17 for total Bcd, 0.5 for Hb. g: Positions $x_{\rm Bcd}$
and $x_{\rm Hb}$ 
as function of embryo length (EL); Bcd: circles, Hb:
squares. Green straight line is $x_{\rm Hb} = 0.49 \times {\rm EL}$.}
\end{figure}

Motivated by recent experimental observations~\cite{Zhu01,Zhao02},
which show that the activity of Bcd itself can be reduced by
interactions with other proteins called co-repressors, we predict the
existence of a second, as yet unidentified, morphogen that diffuses
from a posterior-localized mRNA source. The co-repressor binds to Bcd,
thereby inhibiting its ability to activate {\em hb} expression in the
posterior half of the embryo (see Fig.~\ref{fig:1}a). We show that
such a symmetric activator-co-repressor module can locate the embryo
center reliably. Our model thus generates robust patterns of gene
expression that scale with size, a result of potentially wide
importance. Our calculations reveal that the module is highly robust
against {\em correlated} embryo-to-embryo variations in the synthesis
rates of the anterior activator (Bcd) and posterior
co-repressor. Furthermore, variations in the synthesis rate of Bcd
that are {\em uncorrelated} with that of the co-repressor, are also
filtered far more than in a single-morphogen model. Identifying the
proposed posterior morphogen may require a large-scale search for
maternal mRNA 
localized to the posterior pole, 
which does not correspond to an
already characterized maternal gene.

{\em Mathematical Model.} The outline of our model is shown in 
Fig.~\ref{fig:1}: Bcd and co-repressor protein are injected at
rates $J_B$, $J_R$ at opposite ends of the embryo, where at each time
instant the injection is concentrated at a randomly chosen location $x_B(t)$,
$x_R(t)$  between $5 \mu{\rm m}$ and $30 \mu{\rm m}$ from the respective poles.
 The newly translated proteins diffuse away from the poles with
diffusion constant $D$, and are degraded at a rate $\mu$.  Bcd and the
co-repressor bind with rate $\nu$ to form a complex also with
diffusion constant $D$. We assume that the reverse reaction occurs at
a far slower rate, which we neglect. Only active Bcd (unbound to
co-repressor) binds to the {\em hb} control region on the DNA to
activate its transcription.  Experiments suggest that 
this is a cooperative process~\cite{Ma96}, which we
model with a Hill function with a coefficient of $3$~\cite{Ma96}.
 However, the precise form
of the activation function is not very important: taking other
functional forms or changing the Hill coefficient to unity, for
instance, does not alter our results, as the concentration of active
Bcd is reduced to very low values at mid-embryo due to the
co-repressor (see also below).
Hb also auto-activates its own synthesis via
`synergistic' interactions with (active) Bcd
~\cite{Houchmandzadeh02,SimpsonBrose94}. {\em hb} mRNA has diffusion
constant $D_{hb}$, and is spontaneously degraded with rate
$\mu_{hb}$. Hb protein is manufactured at a rate proportional
to the mRNA concentration, has diffusion constant $D_{\rm Hb}$ and
is spontaneously degraded with rate $\mu_{\rm Hb}$.  These processes
give the following equations, where $[B]$ is the unbound Bcd density,
$[R]$ the co-repressor density, $[C]$ the Bcd-co-repressor
complex density, $[hb]$ the {\em hb} mRNA density and $[{\rm Hb}]$
the density of Hb protein: 
\begin{eqnarray}
\frac{\partial [B]}{\partial t} &=& D \frac{\partial^2[B]}{\partial
  x^2} - \mu [B] - \nu [B] [R] + J_B \delta(x-x_B(t)) \nonumber \\
\frac{\partial [R]}{\partial t} &=& D \frac{\partial^2[R]}{\partial
  x^2} - \mu [R] - \nu [B] [R] + J_R \delta(x-x_R(t)) \nonumber \\
\frac{\partial [C]}{\partial t} &=& D \frac{\partial^2[C]}{\partial
  x^2} - \mu [C] + \nu [B] [R] \nonumber \\ \frac{\partial
  [hb]}{\partial t} &=& D_{hb} \frac{\partial^2[hb]} {\partial x^2} -
\mu_{hb} [hb] + \frac{\beta [B]^3 ( \eta [{\rm Hb}] + \gamma K)}{K^4 +
  [B]^3 (\eta [{\rm Hb}] + K)}\label{eq:hb} \nonumber \\
\frac{\partial [{\rm Hb}]}{\partial t} &=& D_{\rm Hb}
\frac{\partial^2[{\rm Hb}]}{\partial x^2} - \mu_{\rm Hb} [{\rm Hb}] +
\alpha [hb] ~ . \nonumber
\end{eqnarray}
Here, we have used a simplified $1d$ model, where $x$ measures the
distance from the embryo's anterior pole. Zero flux boundary
conditions were imposed at both embryo ends. The wild-type parameters
used were (see also~\cite{Houchmandzadeh02}): EL $ = 460 \mu{\rm m}$, $D = 10 \mu{\rm m}^2
{\rm s}^{-1}$, $D_{hb} = 0.5 \mu{\rm m}^2 {\rm s}^{-1}$, $D_{\rm Hb} =
1 \mu{\rm m}^2 {\rm s}^{-1}$, $J_B = J_R = 0.08 {\rm s}^{-1}$, $\mu =
0.0007 {\rm s}^{-1}$, $\mu_{hb} = 0.004 {\rm s}^{-1}$, $\mu_{\rm Hb} =
0.006 {\rm s}^{-1}$, $\nu = 3 \mu{\rm m} {\rm s}^{-1}$, $\alpha =
0.006 {\rm s}^{-1}$, $\beta = 0.004 \mu{\rm m}^{-1} {\rm s}^{-1}$,
$\eta=1.0$, $\gamma = 0.01$, $K = 0.045 \mu{\rm m}^{-1}$. 
We use the same diffusion
constant/degradation rate for the complex as for unbound Bcd; this 
 ensures that the total Bcd profile decays exponentially
(as seen in experiment~\cite{Houchmandzadeh02}), but is not important
for the basic center finding properties of the model. Spatiotemporal
fluctuations of the components within a single embryo, which could
affect robustness~\cite{England05}, were also modelled but found to be
less important than embryo-to-embryo variability and were therefore
neglected. To simulate the embryo-to-embryo variability, the embryo
length fluctuated by 10\%
~\cite{Houchmandzadeh02}. Other parameters were varied
by imposing a Gaussian distribution about the above average values,
using the following standard deviation divided by the mean: $J_B:
0.05, J_R: 0.05, \mu: 0.3, \mu_{hb}: 0.3, \mu_{\rm Hb}: 0.3, \alpha:
0.3, \beta: 0.3, K: 0.3$. The variations in the protein degradation
rates
were assumed to be correlated, while we studied both
correlated and uncorrelated fluctuations in the Bcd and co-repressor
source strengths. Initially, the system was empty of proteins and {\em
hb} mRNA.  The system reaches a steady state after $\sim 1$ hour,
well before the window in which the relevant experiments were
performed~\cite{Houchmandzadeh02,Spirov03}.
\begin{figure}[t] \centering
\includegraphics[width=8cm]{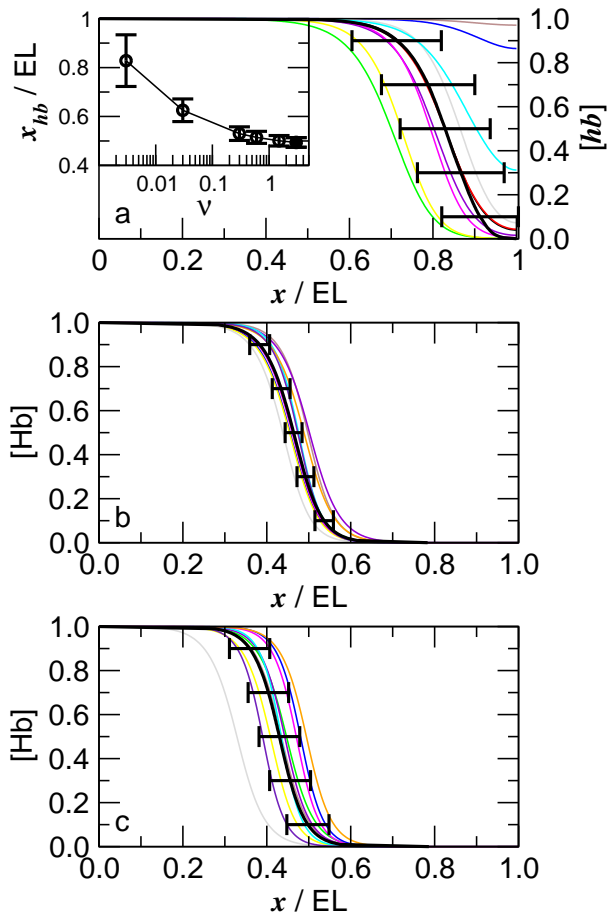} \caption{\label{fig:2} a:
{\it hb} mRNA profiles in simulated Bcd(A52-56) mutant ($\nu=0.003
\mu{\rm m} {\rm s}^{-1}$). Inset: mean/standard deviation for position
of {\em hb} boundary as function of Bcd/co-repressor binding rate
$\nu$ (closed symbol denotes wild type). b: Mean/standard deviation
for Hb profile in mutant with malfunctioning Hb protein. c:
Mean/standard deviation for Hb density profile in simulated {\em
staufen} mutant (see text). Data from 100 simulated embryos; maximum
densities normalized to unity; $10$ typical individual profiles
are also shown.}
\vspace{-0.3cm}
\end{figure}

It is conceivable that an additional activator~\cite{Fu04} diffusing
from the anterior competes with the co-repressor for Bcd
binding. This, however, does not significantly change the
results.  
 Models in
which the co-repressor and/or the Bcd-co-repressor complex can
also bind to the DNA, but cannot then initiate {\em hb} transcription,
again produce very similar results. Schemes in which Bcd and
(co-)repressor {\em only} interact on the DNA, for instance by mutually
excluding each other's DNA binding, provide scaling, but are less
robust to variations in their parameters than the model proposed here.
Furthermore, we have neglected the
posterior morphogen {\em nanos} 
because if {\em nanos} is
removed, the key properties of accuracy and scaling
remain~\cite{Houchmandzadeh02}.  Lastly,
while the co-repressor inactivates Bcd at mid-embryo with
respect to {\em hb} expression, it does not impair Bcd's capacity to
activate other genes in the embryo's posterior, such as {\em giant}
and {\em knirps}~\cite{RiveraPomar95}. Indeed, the Bcd domain that
interacts with the co-repressor 
 plays an inhibitory role in
{\em hb} expression, but a positive role in {\em knirps}
expression~\cite{Fu03}.

An alternative model has been suggested~\cite{AegerterWilmsen05}; it
assumes that the mRNA-localization protein Staufen transports {\em hb}
mRNA away from the poles via a counter-intuitive mechanism relying on
an {\em increased} diffusion rate of {\em hb} mRNA upon binding to
Staufen. However, scaling is found over only 1-2 minutes rather than
the required hour or more. Moreover, it predicts a gradient of
Staufen, which 
is in contradiction with
experiment~\cite{StJohnston91}. Hence, as proposed here, a different
mechanism may be needed.
 
{\em Results.}
Simulation results of the above model, with
uncorrelated fluctuations in the source stengths, are shown in
Fig.~\ref{fig:1}.  The density of Bcd plus Bcd-co-repressor complex
decays exponentially away from the anterior pole (Fig.~\ref{fig:1}b),
in agreement with experiment~\cite{Houchmandzadeh02}. The total Bcd
profile shows considerable embryo-to-embryo variation, with the
characteristic length scale $\lambda = \sqrt{D/\mu}$ of the profile
(Fig.~\ref{fig:1}e) and the position where the density crosses a
threshold (Fig.~\ref{fig:1}f), both showing large fluctuations,
similar to experiment~\cite{Houchmandzadeh02,Spirov03}. However, the
profiles of the co-repressor/unbound Bcd fluctuate far less from
embryo to embryo (Fig~\ref{fig:1}c); the fact that they bind, 
 forces their unbound densities to very low values at
mid-embryo. The active Bcd is then able to precisely activate {\em
hb}: the {\em hb} mRNA (data not shown) and Hb protein
(Fig.~\ref{fig:1}d) densities both pass their half-maximum values at
$x_{\rm Hb}$ with little embryo-to-embryo variability, $x_{\rm
Hb}=0.49 \, (\pm 0.02) \, {\rm EL}$, see also Fig.~\ref{fig:1}f. The
concentration profiles switch between their highest and lowest values
over a length scale of about 0.1 EL. Both the average position and
variation of the Hb boundary, as well as its steepness, are in
agreement with
experiment~\cite{Houchmandzadeh02,Spirov03}. Importantly, the Hb
boundary location scales with the embryo size, unlike the Bcd
profile~\cite{Houchmandzadeh02,Spirov03}
(Fig.~\ref{fig:1}g). Furthermore, scaling is achieved even before the
protein/mRNA concentrations reach steady-state, adding additional
robustness to the model. Finally, scaling can
be obtained even when the source strengths and $\lambda$ are not the
same for Bcd and co-repressor (data not shown).

\begin{figure}[b] \centering
\includegraphics[width=7cm]{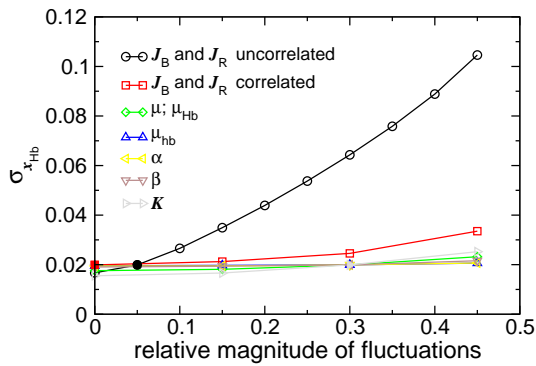} \caption{\label{fig:3} Magnitude of
variations in position of Hb boundary as a function of relative
magnitude (standard deviation $\sigma / {\rm mean~value}$) of
embryo-to-embryo variations in model parameters. Both uncorrelated and
correlated variations in the injection rates $J_B$, $J_R$ are shown
(see text).
When varying the  
fluctuations of one parameter, other
parameters were fixed to their wild type values, indicated by closed
symbols.}
\end{figure}

A critical test of our model comes from varying the {\em bcd} gene
dosage. When the dosage is halved (i.e. Bcd injection rate halved),
the Hb boundary moves to $x_{\rm Hb} = 0.39 \, (\pm 0.02) \, {\rm
EL}$. Simulations with twice and three times normal dosage, but
assuming only 50\% efficiency for the extra {\em bcd} genes as
suggested by experiments~\cite{Houchmandzadeh02,Zhao02} (i.e. $1.5$
and $2$ times the wild type Bcd injection rate), move the boundary to
$x_{\rm Hb} = 0.56 \, (\pm 0.025) \, {\rm EL}$ and $x_{\rm Hb}=0.60 \,
(\pm 0.03) \, {\rm EL}$, respectively.  Both the results on the mean
and the variance agree with experiment~\cite{Houchmandzadeh02}.
Interestingly, the accuracy of {\em hb} expression is only weakly
affected by changes in the {\em bcd} gene dosage: the standard
deviation in $x_{\rm Hb}$ only increases by 50\% when the Bcd
injection rate changes from 0.5 to 2 times the wild type Bcd injection
rate.
More importantly,
the average {\em hb} boundary moves by less than predicted by a simple
threshold model~\cite{Houchmandzadeh02}: the co-repressor gradient
makes the boundary positioning more robust to variations in the
protein production rates. Thus, the posterior morphogen not only
provides scaling, but it also plays an important role as a buffer
against gene expression fluctuations.

Our model is also supported by several mutants. The Bcd(A52-56) mutant
inhibits the ability of Bcd to interact with
co-repressors~\cite{Zhao02}. 
We have modeled this by reducing the Bcd/co-repressor binding by a
 factor of $1000$. Fig.~\ref{fig:2}a shows that this leads to a
 pronounced posterior shift in the {\em hb} expression pattern, in
 agreement with experiment~\cite{Zhao02}.  We also predict that its
 precision will be lost: the {\it hb} boundary fluctuates strongly as
 the posterior co-repressor is no longer able to buffer against Bcd
 fluctuations. 
Experiments have also been performed in mutants that make detectable
but malfunctioning Hb~\cite{Houchmandzadeh02}.  Normally Hb
auto-activates its own
synthesis~\cite{Houchmandzadeh02,SimpsonBrose94}, and our model
includes this effect. If, however, we assume that in this mutant Hb
cannot stimulate its own synthesis (corresponding to $\eta=0$), then
the Hb boundary shifts anteriorly to $x_{\rm Hb}=0.46 \, (\pm 0.02) \,
{\rm EL}$ (Fig.~\ref{fig:2}b), while retaining its scaling
properties. These results all accord with
experiment~\cite{Houchmandzadeh02,SimpsonBrose94}.  Further support
comes from {\em staufen} mutants. Staufen plays a central role in
polar localizing maternal mRNAs~\cite{StJohnston91,Ferrandon94}.
In the $stau^{HL}/stau^{r9}$ mutants, the precision of the Hb boundary
is lost and many of the Hb profiles are shifted
anteriorly~\cite{Houchmandzadeh02}. In contrast, the Bcd profile is not
altered in the $stau^{HL}$ mutants~\cite{Houchmandzadeh02}. Taken
together, these observations strongly suggest that in the
$stau^{HL}/stau^{r9}$ mutants, co-repressor mRNA is no longer properly
localized at the posterior pole. We have studied this scenario by
injecting 
a random fraction of the co-repressor
close to the posterior (as described above), with the remainder
randomly distributed throughout the embryo. The results are shown in
Fig.~\ref{fig:2}c: Hb boundary precision is lost, and many profiles
shift towards the anterior ($x_{\rm Hb}=0.43 \, (\pm 0.05) \, {\rm
EL}$), in agreement with experiment~\cite{Houchmandzadeh02}.

To isolate which embryo-to-embryo variations are most important, we
studied the magnitude of the fluctuations in the Hb boundary position
as a function of the embryo-to-embryo variability of various
parameters. Correlated variations that affect Bcd and the co-repressor
similarly have little effect, due to the symmetry of the
activator-co-repressor geometry (see Fig.~\ref{fig:3}). If, for
instance, both Bcd and co-repressor are degraded by the same pathway,
then the embryo-to-embryo variations in their degradation rates are
correlated and thus filtered out (see Fig.~\ref{fig:3}).  Furthermore,
it is conceivable that not only degradation, but also synthesis of Bcd
and co-repressor exhibit correlated embryo-to-embryo fluctuations,
because they share, for instance, the same components of the
transcription/translation apparatus. These correlated variations of
the Bcd/co-repressor injection rates are again filtered (see
Fig.~\ref{fig:3}).  However, Bcd/co-repressor synthesis may also
exhibit some uncorrelated variation. As we discussed above, our
symmetric activator-co-repressor system buffers fluctuations far more
than would be possible in a single morphogen system. Nevertheless,
these uncorrelated fluctuations are not completely filtered out and,
if large enough, can adversely affect the accurate positioning of the
{\em hb} expression domain (see Fig.~\ref{fig:3}). Our model therefore
predicts that Bcd synthesis must be tightly controlled, at least
relative to that of the co-repressor.
      
We thank J. Ma, D. Bray, H. Bakker and M. Dogterom for comments and The
Isaac Newton Institute for funding. MH acknowledges funding from The
Royal Society. This work is part of the research program of FOM/NWO.

\end{document}